\documentstyle{article} \input{amssym.def} \input{amssym.tex}
\undefine\hbar \newsymbol\ltimes 226E \newsymbol\upharpoonright 1316
 \newsymbol\hbar 207E
\newsymbol\square 1003 \newsymbol\blacksquare 1004
\def\Box{\hbox{\vrule height1ex\kern-0.4pt \vbox to 1ex{\hrule
width1ex\vfil\hrule width1ex}\kern-0.4pt\vrule height1ex}}
\newcommand{\sqr}[2]{{{\vcenter{\vbox{\hrule height.#2pt \hbox{\vrule
width.#2pt height#1pt \kern#1pt \vrule width.#2pt} \hrule
height.#2pt}}}}}

 \newcommand{\til}{\tilde}

\newcommand{\be}{\begin{equation}}  \newcommand{\ee}{\end{equation}}

\newcommand{\dl}{\delta} 
 
 \newcommand{\th}{\theta}

\newcommand{\rh}{\rho} \newcommand{\sg}{\sigma} \newcommand{\ta}{\tau}
 \newcommand{\Ph}{\Phi}
\newcommand{\phv}{\varphi} 
\newcommand{\ps}{\psi} \newcommand{\Ps}{\Psi} 
 
\newcommand{\raw}{\rightarrow} \newcommand{\A}{\frak A}
\newcommand{\B}{\frak B} 
\newcommand{\C}{{\Bbb C}} 
\newcommand{\bib}{\bibitem} \newcommand{\cin}{C^{\infty}}
 
\renewcommand{\H}{\mbox{$\cal H$}}   \newcommand{\n}{\parallel}

 \newcommand{\R}{{\Bbb R}}
   \newcommand{\notp}{p
\kern-.48em /} \newcommand{\ci}{\cite}
\newcommand{\bea}{\begin{eqnarray}} \newcommand{\eea}{\end{eqnarray}}
\newcommand{\ot}{\otimes} \newcommand{\half}{\mbox{\footnotesize
$\frac{1}{2}$}} 
 \newcommand{\PA}{{\cal P}({\frak A})}
\newcommand{\qh}{q_{\hbar}} \newcommand{\cci}{C_c^{\infty}}
\renewcommand{\P}{{\cal P}} \newcommand{\CG}{{\cal G}}
\newcommand{\CA}{{\cal A}} \newcommand{\bg}{{\bf g}}
\newcommand{\g}{{\frak g}} \newcommand{\BT}{\Bbb T}
\newcommand{\la}{\langle} \newcommand{\ra}{\rangle} \topmargin =
0.25in \textheight = 9.0in \textwidth = 6.25in \oddsidemargin =
0.truein \evensidemargin=0.truein \headheight=0pt \headsep=0pt
\begin{document}
\begin{center}
{\Large\bf Constrained quantization in algebraic field
 theory}\vspace{2mm} {\large N.P.~Landsman\footnote{KNAW Fellow; {\em
 npl@wins.uva.nl}}}\\ Department of Mathematics, University of
 Amsterdam\\ Plantage Muidergracht 24, NL-1018 TV AMSTERDAM
\end{center}
{\em Quantization relates Poisson algebras to $C^*$-algebras. The
analysis of local gauge symmetries in algebraic quantum field theory
is approached through the quantization of classical gauge theories,
regarded as constrained dynamical systems. This approach is based on
the analogy between symplectic reduction and Rieffel induction on the
classical and on the quantum side, respectively. Thus one is led to a
description of e.g. $\th$-angles and gauge anomalies in the algebraic
setting. }
\section{Quantization of observables}
The common structure of the algebra of observables of a classical or a
quantum system is that of a Jordan-Lie algebra. This is a real vector
space $\A$ equipped with two bilinear maps $\circ$ and $\{\;,\;\}$,
where $\circ$ is symmetric and $\{\;,\;\}$ is anti-symmetric. For each
$A\in\A$ the map $B\raw \{A,B\}$ must be a derivation of $\circ$ as
well as of $\{\;,\;\}$ (it follows that the latter is a Lie bracket),
and the identity $(A\circ B)\circ C-A\circ (B\circ C)= k\{\{A,C\},B\}$
must hold for some $k\in\R$.  A Poisson algebra is a Jordan-Lie
algebra in which $k=0$ \ci{PTP,TCQ}.

The main sources of Poisson algebras are Poisson manifolds, which are
smooth manifolds $P$ equipped with a Poisson bracket $\{\;,\;\}$ on
$\A=\cin(P)$.  The product $f\circ g$ is then simply given by
pointwise multiplication $fg$.  In quantization theory one often works
with a sup-dense subalgebra $\til{\A}^0$ of $C_0(P)$, rather than with
$\cin(P)$ \ci{TCQ}.

 Jordan-Lie algebras with $k>0$ are obtained by taking $\A$ to be the
self-adjoint part of a $C^*$-algebra $\A_{\C}$. One then takes $A\circ
B=\half(AB+BA)$ and $\{A,B\}=i(AB-BA)/\hbar$ for some
$\hbar\in\R\backslash\{0\}$; note that these maps indeed preserve
self-adjointness (unlike the associative product in $\A_{\C}$).  One
then has $k=\hbar^2/4$.  (Similarly, to get $k<0$ one takes $\A$ to be
an $R^*$-algebra and omits the $i$.)  A Jordan-Lie algebra $\A$ with
$k>0$ is the the self-adjoint part of a $C^*$-algebra iff $\A$ is a
so-called $JB$-algebra in $\circ$, in other words, iff it is a Banach
algebra in which the inequality $\n A\n^2\,\leq\,\n A^2+ B^2\n$ holds
(here $A^2=A\circ A$).

These considerations inspire the definition of a strict quantization
of a Poisson algebra $\A^0$ as a family of $C^*$-algebras
$\{\A_{\C}^{\hbar}\}_{\hbar\in I}$, where
$I\backslash\{0\}\subseteq\R$ has $0$ as an accumulation point, and
for each $\hbar$ one has a linear map $Q_{\hbar}:\til{\A}^0\raw
\A^{\hbar}$ whose image is dense. One naturally requires that for all
$f\in \til{\A}^0$ the function $\hbar\raw \n Q_{\hbar}(f)\n$ is
continuous on $I$ (in particular, $\lim_{\hbar\raw 0} \n
Q_{\hbar}(f)\n =\n f\n_{\infty}$), that $\lim_{\hbar\raw 0} \n
\{Q_{\hbar}(f),Q_{\hbar}(g)\}_{\hbar} -Q_{\hbar}(\{f,g\})\n =0$, and
finally that $ \lim_{\hbar\raw 0} \n Q_{\hbar}(f)\circ
Q_{\hbar}(g)-Q_{\hbar}(f\cdot g)\n =0$.  In the present form the first
two conditions were proposed by Rieffel \ci{Rie89}; a heuristic
version of the last two goes back to Dirac and von Neumann,
respectively. `Deformation' quantization `deforms' the symmetric
product on $\til{\A}^0$; this is equivalent to the present approach
iff $Q_{\hbar}(\til{\A}^0)$ is closed under multiplication and
$Q_{\hbar}(f)=0$ implies $f=0$. Our approach is particularly suitable
to describe what currently is called Berezin-Toeplitz quantization.
\section{Quantization of pure states}
One may look at quantization from the perspective of pure states
rather than algebras. This requires an understanding of the extent to
which the properties of a $C^*$-algebra $\A_{\C}$ are encoded in its
pure state space $\PA$. It turns out that $\PA$ should be seen as a
Poisson space with a transition probability; this is a generalization
of a Poisson manifold which is not a manifold itself (but rather a
uniform space in the sense of point-set topology), but nonetheless is
foliated by symplectic leaves (in the smooth case these are the
``irreducible'' subspaces of a Poisson manifold, which are somewhat
comparable with the primitive ideals of an algebra).  The symplectic
leaves of the pure state space of a $C^*$-algebra are projective
Hilbert spaces equipped with the well-known Fubini-Study symplectic
form.

 A transition probability on a pure state space $\P$ is a function
 $p:\P\times\P\raw [0,1]$ with appropriate properties, such as
 $p(\rh,\sg)=1$ iff $\rh=\sg$. On a classical pure state space one has
 $p(\rh,\sg)=\dl_{\rh\sg}$, whereas on a $C^*$-algebra the function
 $p$ is defined through the inner product on the Hilbert spaces
 defining the symplectic leaves (which are orthogonal to each other)
 in the well-known way.  Roughly speaking, the Poisson structure on
 $\PA$ encodes the commutator on $\A$, whereas the transition
 probability encodes the Jordan product $\circ$; in conjunction these
 define the associative product on $\A_{\C}$. The norm on $\A$ is
 related to the uniform structure on $\PA$. See \ci{PTP,TCQ}.

The `dual' picture of quantization is now as follows \ci{TCQ}: one has
a family of injections $q_{\hbar}$ of a Poisson manifold $P$ into the
pure state space ${\cal P}(\A^{\hbar})$ of a $C^*$-algebra
$\A^{\hbar}_{\C}$, such that $\lim_{\hbar\raw 0}
p(\qh(\rh),\qh(\sg))=\dl_{\rh\sg}$. In other words, the
quantum-mechanical transition probabilities should converge to the
classical ones. Moreover, the $\qh$ should relate the Poisson
structure on $P$ to that on each ${\cal P}(\A^{\hbar})$ as
appropriate.

Under suitable conditions the quantization $\qh$ of a (locally
compact) symplectic manifold $P=S$ leads to a strict quantization
$Q_{\hbar}$ of $\til{\A}^0=\cci(S)$ into $\A^{\hbar}$ \ci{TCQ}: one
defines $Q_{\hbar}(f)\in \A^{\hbar}$ as the operator whose expectation
value in a pure state $\ps$ on $\A^{\hbar}$ equals
$$
\ps(Q_{\hbar}(f))= \int_S d\mu^{\hbar}(\sg)\, p(\qh(\sg),\ps) f(\sg).
$$ 
Here $\mu^{\hbar}$ is a measure on $S$ which is usually proportional
to the Liouville measure $\mu_L$; the proportionalilty constant blows
up as $\hbar\raw 0$.  In this case $Q_{\hbar}$ is a completely
positive map, and the Stinespring theorem assigns a natural role to
the Hilbert space $L^2(S,\mu_L)$ that always pops up in geometric
quantization and in the Koopman approach to classical mechanics.  For
example, Berezin-Toeplitz quantization is of this sort.

It should be emphasized that the approach advertised here is not a
technique for obtaining quantum mechanics from classical methods (such
as geometric quantization); it is rather the study of the
correspondence between intrinsically defined classical and quantum
theories. Even if one cannot relate a given Poisson algebra (or
Poisson manifold) to a $C^*$-algebra (or its pure state space) by
quantization, one may seek constructions and theorems in both settings
that are `morally' equivalent. This idea plays an essential role in
what follows.
\section{Classical reduction}
Classical gauge theories may be described in terms of symplectic
reduction \ci{Arms}, in that the phase space $S^p$ of physical degrees
of freedom may be obtained from its unphysical counterpart $S$ as an
infinite-dimensional Marsden-Weinstein quotient \ci{AM} with respect
to the gauge group $\CG$ (whose Lie algebra is denoted by $\g$). This
reduction procedure may be carried out covariantly; for simplicity we
work in the temporal gauge $A_0=0$. In that case $S$ is an appropriate
space of spatial gauge fields $A$ and their conjugate momenta $E$.
Let $J:S\raw\g^*$ be the momentum map defined by the action of $\CG$
on $S$.  The constraint $J=0$ is Gauss' law, so that the reduced phase
space $J^{-1}(0)/\CG$, in which gauge-equivalent field configurations
are identified, indeed equals $S^p$.

From the perspective of the algebra of observables the reduction
procedure looks as follows: one defines the Poisson algebra of weak
observables $\A^p_w$ as the set of $\CG$-invariant functions $f$ in
$\cin(S)$ (where $f\in\A_w^p$), and subsequently constructs the
physical observable $\pi^p(f)$ in $\cin(S^p)$ as the restriction of
$f$ to $J^{-1}(0)$. The Poisson algebra of classical physical
observables is $\A=\pi^p(\A_w^p)$, which is evidently a quotient of
$\A_w^p$.

As a simple illustration we consider Yang-Mills theory on a circle
$\BT$ with structure group $G$ (whose Lie algebra we denote by
$\bg$). In that case the gauge group $\CG$ is a loop group on $G$, and
the appropriate choice is to define $\CG$ to consist of the Sobolev
space of all continuous loops with finite energy. In other words,
$\CG=H_1(\BT,G)$.  This forces the space $\CA$ of all connections to
be the real Hilbert space $L^2(\BT,\bg)$, so that $S=T^*\CA\simeq
L^2(\BT,\bg_{\C})$. One can solve Gauss' law on the circle in terms of
Wilson loops, and the reduced phase space comes out to be
$T^*(G/Ad(G))$. This is a finite-dimensional space! See \ci{LW} for
details and references to the literature on this fashionable topic.

To see how this is to be quantized it turns out to be helpful to
 generalize the Marsden-Weinstein reduction procedure \ci{RMW}.
 Instead of $\g^*$ we consider an arbitrary Poisson manifold $P$, with
 realization $S$, i.e., we suppose an (anti) Poisson morphism $J:S\raw
 P$ is given (the `generalized momentum map'), with $S$ symplectic.
 Let a second realization $\rh:S_{\rh}\raw P$ be given. We then form
 the fiber product $C=S*_P S_{\rh}=\{(x,y)\in S\times
 S_{\rh}|J(x)=J_{\rh}(y)\}$.  The reduced space is then defined as
 $S^{\rh}=(S*_P S_{\rh})/{\cal F}_0$, where ${\cal F}_0$ is the null
 foliation of the induced pre-symplectic form on $S*_P S_{\rh}$.

When $\CG$ is connected and $J$ is the usual momentum map, this
construction specializes to Marsden-Weinstein reduction if we choose
$S_{\rh}=\{0\}$ (or, more generally, as an arbitrary coadjoint orbit
in $\g^*$) and $\rh$ as the inclusion map. One then has $\pi^0=\pi^p$
as defined earlier. The case where $\CG$ is disconnected will be
discussed later on.

Back to the general case, let $\A_w$ be the Poisson subalgebra of
$C^{\infty}(S)$ consisting of all functions $f$ with the property that
for all $g\in C^{\infty}(P)$ the Poisson bracket $\{J^*g,f\}$ vanishes
at all points of $S$ which are relevant to $C$. We can define an
`induced representation' $\pi^{\rh}$ of $\A_w$ on $S^{\rh}$ by
$\pi^{\rh}(f)([x,y])=f(x)$, where $[x,y]\in S^{\rh}$ is the image of
$(x,y)\in C$ under the canonical projection from $C$ to $C/{\cal
F}_0$. By our definition of $\A_w$ this is well-defined, and the image
$\pi^{\rh}(\A_w)$ is a quotient of $\A_w$.
\section{Quantum induction}
Traditional constrained quantization (Dirac) quantizes $S$ into a
Hilbert space $\H$, and subsequently attempts to define the Hilbert
space of physical states as a subspace of $\H$ on which the quantized
constraints hold. This leads to endless problems. Instead, we propose
a quantum version of the classical reduction procedure in the previous
section as a whole. The central idea is to `quantize' the
unconstrained system in such a way that the presence of constraints is
taken into account, without assigning a specific value to them.

The mathematical implementation of this idea uses the concept of a
(Hilbert) $C^*$-module. This is a complex linear space $L$ equipped
with a right-action of a $C^*$-algebra $\B$, as well as a $\B$-valued
inner product $\la\;,\;\ra_{\B}$ which is positive (i.e.,
$\la\ps,\ps\ra_{\B}\geq 0$ in $\B$) and $\B$-equivariant in that $\la
\phv,\ps B\ra_{\B}=\la \phv,\ps\ra_{\B}B$. Also, $L$ must be closed in
the norm $\n\ps\n^2=\n\la\ps,\ps\ra_{\B}\n$. Assuming that $\B$ is the
quantization of a suitable subalgebra of $\cin(P)$ (cf.\ section 1),
the right-action of $\B$ on $L$ is to be thought of as the
quantization of the pull-back $J^*:\cin(P)\raw\cin(S)$ of the
generalized momentum map $J:S\raw P$, whereas $\la\;,\;\ra_{\B}$ is
the quantization of $J$ itself \ci{RMW}.

The quantum analogue of symplectic reduction is then given by Rieffel
induction \ci{Rie74}. This procedure starts from a represenation
$\pi_{\rh}(\B)$ on a Hilbert space $\H_{\rh}$ (with inner product
$(\;,\;)_{\rh}$), which is the quantum counterpart of the pull-back of
the realization $\rh:S_{\rh}\raw P$ (choosing $\pi_{\rh}$ assigns a
value to the constraints in the appropriate quantum-mechanical
sense). This leads to a sesquilinear form $(\;,\;)_0$ on
$L\otimes\H_{\rh}$, defined by linear extension of $ (\ps\otimes
v,\phv\otimes w)_0=(\pi_{\rh}(\la \phv,\ps\ra_{\B})v,w)_{\rh}$.  This
form is positive semi-definite, with null space ${\cal N}$; the
induced space $\H^{\rh}$ is simply defined as the completion of
$L\otimes\H_{\rh}/{\cal N}$ in the inner product inherited from
$(\;,\;)_0$.

The step of quotienting by $\cal N$ is evidently the quantum analogue
of the passage from the classical constraint hypersurface $C$ to the
physical phase space $C/{\cal F}_0$. However, the set $C$ on which the
classical constraints hold has no quantum counterpart; in quantum
mechanics it is not necessary to solve the constraints. This is a
crucial difference between the present approach and traditional ones
such as that of Dirac (who, to give due credit, did recognize that
only one of the two steps of classical reduction needed to be
quantized; unfortunately, he picked the wrong one).

The algebra $\A_w$ of weak quantum observables of the constrained
system consists of those operators $A$ on $L$ which are self-adjoint
with respect to $(\, , \,)_0$, i.e., $(A\Ps,\Ph)_0=(\Ps,A\Ph)_0$ for
all $\Ps,\Ph\in L\ot\H_{\rh}$ (here $A$ is identified with $A\ot {\Bbb
I}$).  For such operators $A{\cal N}\subseteq {\cal N}$, so that the
quotient action $\pi^{\rh}(A)$ on $L\otimes\H_{\rh}/{\cal N}$ is
well-defined. This `induced representation' may be extended to
$\H^{\rh}$ when a suitable boundedness assumption is satisfied. The
quantum algebra of physical observables $\pi^{\rh}(\A_w)$ is then a
quotient of the algebra $\A_w$ of weak observables.

The idea that Rieffel induction is the quantum analogue of symplectic
reduction is supported by the fact that the main theorems on induced
representations (the theorem on induction in stages and the
imprimitivity theorem \ci{Rie74}) have analogues for symplectic
reduction, whose proofs may almost be copied from the `quantum' case
\ci{RMW,TCQ} (also see \ci{Xu} for a different approach to the
symplectic imprimitivity theorem).
\section{Gauge theory on a circle}
The classical Marsden-Weinstein quotient $S^0=J^{-1}(0)/\CG$ with
finite-dimensional $\CG$ may be quantized according to the above
scheme. One starts by quantizing the unconstrained phase space $S$
with its given $\CG$-action by a Hilbert space $\H$ carrying a unitary
representation $U$ of $\CG$, and subsequently tries to find a dense
subspace $L_0\subseteq\H$ such that the function $x\raw (U(x)\Ps,\Ph)$
is in $L^1(G)$ for all $\Ps,\Ph\in L_0$. When $G$ is compact one may
choose $L_0=\H$.  With $\B=C^*(G)$ the $\B$-valued inner product on
$L_0$ is defined by $\la \Ps,\Ph\ra_{C^*(G)}=(U(x)\Ph,\Ps)$. Under
suitable conditions this may be closed into a $C^*$-module $L$; note
that in general $L\neq\H$. There is a bijective correspondence between
unitary representations $U_{\rh}$ of $G$ and representations
$\pi_{\rh}$ of the group algebra $C^*(G)$, given (initially on
$L^1(G)$) by $\pi^{\rh}(f)=\int_G dx f(x)U_{\rh}(x)$. Hence the form
$(\;,\;)_0$ on $L\otimes\H_{\rh}$ comes out as the group average
$$
  (\Psi,\Phi)_0=\int_{\CG}dx\,(U(x)\ot U_{\rh}(x)\Psi,\Phi) . $$

One may start the induction procedure from this expression, forgetting
$C^*(G)$ and the derivation of the formula. This philosophy is
necessary in applying the technique to gauge theories, where the gauge
group $\CG$ is not locally compact and fails to possess a Haar measure
$dx$, so that $C^*(G)$ is not defined. We illustrate this for a gauge
theory on a circle \ci{LW}, as discussed in section 3 (also see
\ci{LWi} for abelian gauge theories on $\R^3$).

 For the Hilbert space $\H$ of the unconstrained system we take the
bosonic Fock space $\exp(S)$ over the classical phase space
$S=L^2(\BT,\g_{\C})$, and for $U(\CG)$ we take the well-known `energy
representation' \ci{AHK}.  The domain $L_0$ is the linear span of the
exponential (or `coherent') vectors in $\exp(S)$; these are of the
form $\sqrt{\exp}(A)=\sum_{n=0}^{\infty}\otimes^n A/\sqrt{n!}$. It
turns out that, heuristically speaking, the would-be Haar measure on
$\CG$ combines with a Gaussian factor in the matrix element of $U(x)$
to form the Wiener measure $\mu_W$ (cf.\ its emergence in the
Feynman-Kac formula). Since $\CG$ has $\mu_W$-measure zero, the domain
of integration must be extended from $\CG$ to its sup-closure
$C(\BT,G)$ (the fact that $U$ cannot be extended beyond $\CG$ is
inconsequential, since $(\Psi,\Phi)_0$ remains well-defined).

Since the physical phase space of the classical theory is obtained by
reduction from 0, it seems natural to define its quantum counterpart
by inducing from the trivial representation $U_0(\CG)=1$ on $\H_0=\C$
(so that $L\otimes\H_{\rh}\simeq L$). The resulting Wiener integral
over $C(\BT,G)$ can be performed exactly, so that the explicit form of
the induced space $\H^0$ can be derived. This is most easily done by
guessing what $\H^0$ should be (in the present case heuristic
approaches suggest that $\H_{\rm guess}^0=L^2(G/Ad(G))$), and then
confirming this guess by constructing a map $V:L\raw \H_{\rm guess}^0$
satisfying $(V\Ps,V\Ph)=(\Ps,\Ph)_0$.  For when $\H_{\rm guess}^0$
indeed equals $\H^0$, the map $V$ is simply the canonical projection,
in terms of which $\pi^0(A)V=VA$.

Interestingly \ci{KKW}, if one reduces by the group of based gauge
transformations, so that $S^p=T^*G$ and $\H^0=L^2(G)$, the image of an
exponential vector $\sqrt{\exp}(A)$ is always a coherent state in
$L^2(G)$ in the sense of Hall \ci{Hall}. This fact allows for an
efficient computation of the induced representation $\pi^0$.
\section{$\th$-angles and anomalies}
The preceding section may be summarized by saying that the algebra of
observables is constructed by inducing from the trivial representation
of the gauge group.  This was justified by the corresponding
construction of the classical phase space by reduction from the zero
level of the momentum map.  In case that the gauge group $\CG$ is
disconnected, however, there is no good reason why one should not
induce from a unitary representation $U_{\th}=\til{U}_{\th}\circ \ta$,
where $\til{U}_{\th}$ is a unitary representation of $\CG/\CG_0$ and
$\ta:\CG\raw \CG/\CG_0$ is the canonical projection (here $\CG_0$ is
the component of $\CG$ containing the unit element).

The origin of this freedom lies in the fact that the quotient $S/D$ of
a symplectic manifold $S$ by a discrete group $D$ is symplectic,
unlike the quotient by a Lie group of dimension $>0$. Hence there is
no momentum map $J$ and no $``0''$ in $J^{-1}(0)$. In other words,
there is no constraint in the classical theory.  To apply this
argument to gauge theories one reduces and induces in stages, firstly
with $\CG_0$ and secondly with the discrete group $\CG/\CG_0$.  We
conclude that in the present approach (generalized) $\th$-angles
emerge from the freedom to induce from $\til{U}_{\th}(\CG)$ rather
than from the trivial representation.

Possible gauge anomalies appear when $U$ is merely a projective
representation of $\CG$.


\begin{thebibliography}{99}
\bib{PTP} N.P. Landsman, {\em Rev.\ Math.\ Phys.} {\bf 9} (1997)
29-57.  \bib{TCQ} N.P. Landsman, {\em Tractatus
Classico-Quantummechanicus}, Springer, New York, 1998.  \bib{Rie89}
M.A. Rieffel, {\em Commun. Math. Phys.} {\bf 122} (1989) 531-562.
\bib{Arms} J.M. Arms, {\em Acta Phys.\ Polon} {\bf B17} (1986)
499-523.  \bib{AM} R. Abraham and J.E. Marsden, {\em Foundations of
Mechanics}, 2nd ed., Addison Wesley, Redwood City, 1985.  \bib{LW}
N.P. Landsman and K.K. Wren, {\em Nucl.\ Phys.} {\bf B}, in press
({\em hep-th/9706178}).  \bib{RMW} N.P. Landsman, {\em J.\ Geom.\
Phys.} {\bf 15} (1995) 285-319.  \bib{Rie74} M.A. Rieffel, {\em Adv.\
Math.} {\bf 13} (1974) 176-257.  \bib{Xu} P. Xu, {\em Commun.\ Math.\
Phys.} {\bf 142} (1991) 493-509.  \bib{LWi} N.P. Landsman and U.A.
Wiedemann, {\em Rev.\ Math.\ Phys.} {\bf 7} (1995) 923-958.
\bibitem{AHK} S. Albeverio and R. Hoegh-Krohn, {\em Comp.\ Math.}
{\bf 36} (1978) 37-52.  \bib{KKW} K.K. Wren, Constrained quantization
and $\th$-angles II.  \bib{Hall} B.C. Hall, {\em J.\ Funct.\ Anal.}
{\bf 122} (1994) 103-151.  \end{thebibliography}
\end{document}